\documentclass[preprint,aps,prl]{revtex4}
\usepackage{epsfig,graphics}
\newcommand{\bea}{\begin{eqnarray}}
\newcommand{\eea}{\end{eqnarray}}
\newcommand{\be}{\begin{equation}}
\newcommand{\ee}{\end{equation}}

\newcommand{\btof}{\gamma_{b \rightarrow f}}
\newcommand{\ftob}{\gamma_{f \rightarrow b}}

\begin{document}
\title{Self-organized Pattern Formation in Motor-Microtubule Mixtures}

\author{Sumithra Sankararaman\footnote{E-mail : sumithra@imsc.res.in}
and Gautam I. Menon\footnote{E-mail : menon@imsc.res.in}}
\affiliation{The Institute of Mathematical Sciences,\\ C.I.T. Campus,
Taramani, Chennai 600 113,\\ India.}

\author{P.B. Sunil Kumar\footnote{E-mail : sunil@physics.iitm.ac.in}}
\affiliation{Department of Physics,\\ Indian Institute of Technology
Madras, \\ Chennai 600 036,\\ India.}

\begin{abstract}
We propose and study a hydrodynamic model for pattern
formation in mixtures of molecular motors and
microtubules. The steady state
patterns we obtain in different regimes of parameter space
include arrangements of vortices and asters separately as
well as aster-vortex mixtures and fully disordered states. Such 
stable steady states are observed in experiments {\it in vitro}.
The sequence of patterns obtained in the experiments can be
associated with smooth trajectories in a non-equilibrium
phase diagram for our model.
\end{abstract}
\date{\today}
\pagebreak
\maketitle

The mitotic spindle is a remarkable self-organized
structure formed when a eukaryotic cell divides\cite{brucealberts}.  
It consists of two separated inter-penetrating radial arrays
(asters) of long, semi-flexible polymeric filaments with
polar character, called
microtubules.  Molecular motors such as kinesins and
dyneins are protein molecules which ``walk'' unidirectionally along
microtubules and exert forces on them, while consuming
energy derived from the hydrolysis of ATP.  Motor-microtubule
interactions play a fundamental role in the formation 
of the spindle\cite{brucealberts}.

Interestingly, cell fragments containing both
motors and microtubules have been found to
exhibit self-organized aster-like 
structures\cite{rodionovborisy,heald}.
Asters and vortices are seen {\it in vitro}, in
experiments on mixtures of molecular motors,
microtubules and ATP in a confined
quasi-two-dimensional geometry
\cite{nedelecsurreymaggsleibler,surreynedelecleiblerkarsenti}.
Studying pattern formation in such simplified contexts may
be useful in understanding the vastly more
complex problem of spindle 
formation\cite{karsentivernos,nedelecsurreykarsenti,nedelecjcb}.

These patterns form at large densities of motors and
microtubules, rendering first-principles molecular
simulations unviable.  Such simulations require
{\it ab-initio} modelling of interaction potentials;
small errors in these potentials could be amplified at
larger scales, thus changing the behavior
fundamentally. Further, cellular pattern formation
occurs under non-equilibrium conditions.  These
considerations motivate hydrodynamic approaches in
which molecular scale information enters only in terms
of coefficients in coarse-grained equations of motion
for a few relevant fields\cite{tonertu,liverpool}.

An early attempt at describing such pattern formation
is due to Lee and Kardar\cite{leekardar} and motivates the approach
described here. Also, detailed Brownian dynamics 
simulations on a simplified model for motors, microtubules
and their interactions have been performed by
N\'ed\'elec and collaborators, yielding valuable
insights\cite{nedelecsurreymaggsleibler,
surreynedelecleiblerkarsenti,nedelecjcb,nedelecsurrey}. 
However, these approaches are 
unsatisfactory in some respects.  The Lee-Kardar model 
fails to reproduce the
complex patterns (arrays of asters, vortices and
spirals) seen in experiments where the effects of
boundaries are minimal. Further, it predicts that
the unique steady state at high motor densities
should be a single vortex whereas experiments 
obtain a lattice of asters in this 
limit\cite{leekardar,notearray,lee}.  The simulations
of N\'ed\'elec {\it et. al.} are computationally
intensive, requiring many parameters to be specified\cite{nedelecjcb}. 
Several of these can be varied substantially without
affecting the patterns formed
-- it is unclear which of them are crucial to pattern
formation and which others play a secondary role.

In this Letter, we propose a theory of pattern formation
in mixtures of molecular motors and microtubules. We
motivate, using symmetry arguments,
hydrodynamic equations of motion for a coarse-grained
2-dimensional vector field (${\bf T}$) representing the
local orientation of microtubules as well as for local motor
density fields $m_f$ and $m_b$ governing densities of
``free'' and ``bound'' motors. Motors which move on
microtubules are called bound motors, whereas those which
diffuse in the solvent are referred to as free motors.
Bound and free motors interconvert at rates
$\ftob$ and $\btof$.
Our equations describe the orientation of microtubules by
complexes of bound motors, resulting in the formation of
patterns at large scales.

Before describing our model we briefly summarize
our results:  Our model generates virtually all
the phases seen in the experiments, including a
lattice of asters, a lattice of
vortices/spirals\cite{notespiral}, a mixture of
asters and vortices as well as fully disordered
phases\cite{notebundle}. We present a
non-equilibrium ``phase diagram'' for our model;
smooth trajectories in this phase diagram can be
related to the sequence of states obtained 
experimentally as the motor density is increased.
Such trajectories rationalize the
difference in the sequence of patterns formed by
different motor species -- such as conventional
kinesins and NCD's.  We also predict the decay of
bound and free motor densities in aster and
vortex configurations.

Our equations are derived in
the following way: In the absence of interconversion terms
changing a bound motor to a free motor, $m_b$ obeys a
continuity equation involving the current of motors
transported along the 
microtubules\cite{leekardar,lee,nedelecsurreymaggs}. The free motor field
$m_f$ obeys a diffusion equation with
a diffusion constant $D$. These fields are coupled
through interconversion terms $\ftob$ (free $\rightarrow$ 
bound) and $\btof$ (bound $\rightarrow$ free)\cite{nedelecsurreymaggs}.
We ignore spatial variations in the density of
microtubules, concentrating on their orientational 
degrees of freedom. Our equation for ${\bf T}$
includes terms which
reflect the alignment of microtubules by 
bound motors\cite{leekardar}. We also allow for a motor-independent 
stiffness against distortions.  Finally, and most crucially for
our purposes, symmetry permits 
a term of linear order in gradients of the bound motor density 
field to appear in the equation of motion
for the microtubule field. Putting these ingredients together, we
obtain
\bea 
\partial_t m_f &=& D \nabla^2 m_f - \ftob m_f + \btof m_b \label{free_den}\\
\partial_t m_b &=& -A \nabla.(m_b {\bf T}) + \ftob m_f -\btof m_b \label{bound_den}\\
\partial_t {\bf T} &=& {\bf T}(\alpha - \beta T^2) + \gamma m_b \nabla^2 
{\bf T} +  \gamma^{\prime} \nabla m_b . \nabla {\bf T}  
+ \kappa\nabla^2 {\bf T}+S\nabla m_b 
\eea
The first term in the equation for 
$m_b$ describes the motion, with velocity $A$, of bound motors
along microtubules. 
While $\ftob$ should, in principle, depend on the
local density of microtubule\cite{lee,nedelecsurreymaggs}, we 
argue that such density variations can be {\em neglected} following 
coarse-graining
to the mesoscopic length-scales of relevance to this
problem and at the relatively large concentrations of 
microtubules in the experiments.  We work in two dimensions 
throughout and 
scale length in units of $D/A \sqrt{\beta/\alpha} $, time in units
of $\beta D/(\alpha A^2)$, motor densities in units of $D/\gamma$ and
the tubule density in units of $\sqrt{\alpha/\beta}$.
We choose $\gamma^{\prime} = \epsilon \gamma$ and vary the
parameter $\epsilon$.

The term in $S$ can be thought of as arising from
the variation of a free-energy-like term 
of the following kind:
$F \sim \int d^2 x {({(\nabla. {\bf T})} + S f(m_b))}^2$.
At non-zero $S$, such a term clearly favors a
spontaneous divergence in the tubule configuration.
When motor complexes bind to two parallel tubules,
they pull these tubules inwards,
generating a wedge-like configuration. Such microscopic 
dynamics is incorporated in simulations\cite{nedelecsurrey} which
obtain a lattice of asters. This term arises purely from
bound motors and should, for small $m_b$, have a strength
linear in $m_b$ {\it i.e.} $f(m_b) \sim m_b$.

Three principal
parameters enter our scaled equations:
$\epsilon$, $S$ and the total motor density $m = m_f + m_b$.
(We take $\ftob = \btof = 0.5$ throughout and
choose $\kappa$ to be $10 \%$ of $m$ so that the
motor-induced stiffness is dominant.)
We solve the scaled system of equations numerically
on an $L \times L$ square grid\cite{numrec}.
The ${\bf T}$ equation is differenced
through the Alternate Direction Implicit (ADI) operator
splitting method in the Crank-Nicholson scheme while the
equations of motion for the motor density are differenced
through an Euler scheme.  The grid spacing is $\delta x =
\delta y = \delta = 1.0$, while the time step is 
$0.1$.  We evolve the equations for typically $2 \times 10^4$
time steps, checking for convergence to steady state.  The
conservation of total number of motors is imposed by the
boundary condition that no motor current 
flows into or out of the system. We choose ``reflecting'' boundary
conditions for the ${\bf T}$ field, with all tubules at the
boundary pointing along the inward normal\cite{notebc}.
We work with systems of several sizes, ranging
from $L=30$ to $L=100$, varying total motor densities in
the range 0.01 to 0.5 in dimensionless units\cite{notenoise}.

Figures 1(a) -- (d) depict four stable configurations
obtained in different regimes of parameter space for an
$L=100$ lattice. 
Fig. 1(a) shows a disordered arrangement of 
microtubules obtained at very low motor densities ($m = 0.005$)
with $\epsilon = 0.5$ and $S = 0$.
Figure 1(b) shows an aster-vortex mixture obtained at
$m = 0.01$ at the same values of $\epsilon$
and $S$. Note the presence both of well-formed
asters and of vortices in the configurations. 
This figure is to be contrasted to Fig. 1(c), obtained at
$m = 0.05$, taking $\epsilon = 5$ and $S=0.001$. Note the
the absence of asters in this regime of parameter space. Finally, 
Fig 1(d), obtained with $m = 0.5$,
$\epsilon = 1$ and $S=1$, illustrates a lattice of asters
phase, with asters being the only stable defects present. We can
vary the sizes and numbers of asters obtained in configurations 
such as the one shown in Fig. 1(d), by changing $S$. A larger
$S$ yields a large number of small asters, while smaller
values of $S$ yield a smaller number of large 
asters\cite{nedelecsurreymaggs}.

We also predict complex motor density
profiles in aster configurations\cite{notevortex}. Our 
results here differ
both from those of Lee and Kardar
as well as those of N\'ed\'elec, Surrey
and Maggs (NSM)\cite{nedelecsurreymaggs}.
Assuming steady state aster configurations 
${\bf T} = -{\bf \hat r}$
and solving the time-independent equations in steady state
yields free motor densities
which are a combination of confluent hypergeometric 
Kummer functions\cite{unpublished}.
The bound motor density field satisfies
$\partial_r m_f(r)  = - m_b(r)$. The 
solution obeys the appropriate boundary conditions 
(no current at the boundary) and satisfies the normalization 
constraint on $m$.  We obtain density profiles for free 
motors as
\bea
m_f(r)  &\sim & \frac{e^{-r/\xi}}{r^p} 
\eea
where $p = \frac{1}{2}(1 - \frac{\btof}{\sqrt{\btof^2+4\ftob}})$,
and the inverse of the ``decay length'' $\xi$ is defined as
$\xi^{-1} = \Big|\frac{(\btof - \sqrt{(\btof^2+4\ftob})}{2}\Big| = 
\Big|\frac{p\btof}{2p-1}\Big|$.
The decay length $\xi$ and the exponent $p$ depend on
$\ftob$ and $\btof$.  

Provided $\xi$ is large compared to the scale of the
aster (we estimate $\xi > 40 \mu m$ for typical
parameter values), the decay of motor
densities is governed principally by the power-law
$m_f \sim 1/r^p$. With $\ftob = \btof = 0.5$,
we obtain $p = 1/3$, although we can generate, as
$\ftob$ and $\btof$ are varied, exponents which vary
continuously in the interval [0:0.5].  (Power-law
decays of $m_f$ are also obtained in the NSM model
Ref.\cite{nedelecsurreymaggs}.)

We relate our scaled parameters to typical experimental
values in the following way: The tubule density, scaled in
terms of $\sqrt{\alpha/\beta}$, is chosen to be unity. The
diffusion constant $D$ is about $20 {\mu m}^2/s$ and $A
\sim 1 \mu m/s$, defining basic units of length and time
as $20 \mu m$ and $20$ seconds respectively. A tubule
density of $1$ implies that over a coarse-graining length
of $400 {\mu m}^2$, there are around $400$ microtubules, a
value close to that used in the simulations
\cite{surreynedelecleiblerkarsenti}.  
The physical length of our simulation box is about $200 \mu
m$, which is in the experimentally relevant regime.  
Our choice for $\ftob$ and $\btof$
corresponds to physical rates of
$0.005 s^{-1}$ to $0.05 s^{-1}$, slightly smaller than
those in the simulations\cite{nedelecjcb}; using 
larger rates does not affect our conclusions here.

Our results are summarized in the qualitative phase diagram
of Fig. 2 which shows phases in the
three-dimensional space spanned by $\epsilon$,
$S$ and $m$.
For $S=0$, we obtain a
disordered phase at low motor density, which
becomes a lattice of vortices at somewhat higher
motor densities. Large values of $\epsilon$
($\epsilon \geq 1$) yield well-formed
vortex-like configurations while small 
$\epsilon$ yields structures better
described as aster-vortex mixture states.  At
intermediate values of $\epsilon$ and $m$, spirals 
rather than vortices appear to dominate. At large $m$,
with $S=0$ and large $\epsilon$, a single vortex is 
obtained\cite{leekardar}.

For non-zero but small $S$, these phases
appear to continue out of the $S=0$ plane but are
rapidly replaced by a lattice of asters phase for
larger $S$. A cut of the phase diagram of Fig.
2 at finite $S$ yields two ``phases'' - a
disordered phase at small $m$ and a
lattice of asters at larger $m$. We can thus
understand the sequence of patterns formed upon
increasing $m$ in mixtures
of kinesin constructs with microtubules in terms of 
a trajectory which begins in the $S=0$ (or $S$ sufficiently
small) plane in the disordered phase and transits between
the aster-vortex mixture and lattice of vortices phases
(both of which lie in this plane) as $m$ is increased. As 
$m$ increases further and the effects of the $S$ term
become important, such a trajectory moves out towards 
non-zero $S$, encountering the lattice of asters phase. 

We have also examined the effects of changing the
motor processivity, a quantity proportional to the 
ratio of $\ftob$
to $\btof$. Smaller values of this ratio are
appropriate to molecular motors such as NCD.
At $\ftob = 0.005, \btof =
0.05$, we find that the disordered regime in the phase
diagram of Fig. 2 expands, so that at equivalent values of $m$ 
the disordered phase occupies much of the domain
associated previously with the lattice of vortices. 
Whereas kinesins follow the
sequence {\em disordered -- lattice of vortices --
aster vortex mixture -- lattice of asters} as the
$m$ is increased, a mixture of microtubules
with NCD motors bypasses the lattice of
vortices phase altogether, transiting directly from
the disordered phase to the lattice of 
asters phase in the experiments\cite{surreynedelecleiblerkarsenti}.  In
terms of Fig. 2, the expanded regime of
disordered phase for the NCD motor suggests
that phases such as the lattice of vortices and the 
aster-vortex mixture may be inaccessible at the motor 
densities at which the experiments are done, since 
the effects of the $S$ term might be expected to dominate 
at large $m$.

In conclusion, we have proposed a hydrodynamic theory
capable of generating the sequence of complex phases
seen in experiments on pattern formation in mixtures of
molecular motors and microtubules and constructed a
non-equilibrium phase diagram which contains these phases.
The sequence of
phases seen in the experiments on varying motor densities
can be reproduced in terms of smooth trajectories in this
phase diagram.  Our results are also predictive, since we
can argue that smooth trajectories in parameter space
cannot connect disjoint phases directly. Further
details will appear elsewhere\cite{unpublished}.

We thank G. Date, J. Samuel, M. Rao and Y. Hatwalne 
for useful discussions.  PBS thanks CSIR, India for support.

\newpage

\begin{figure}
\epsfbox{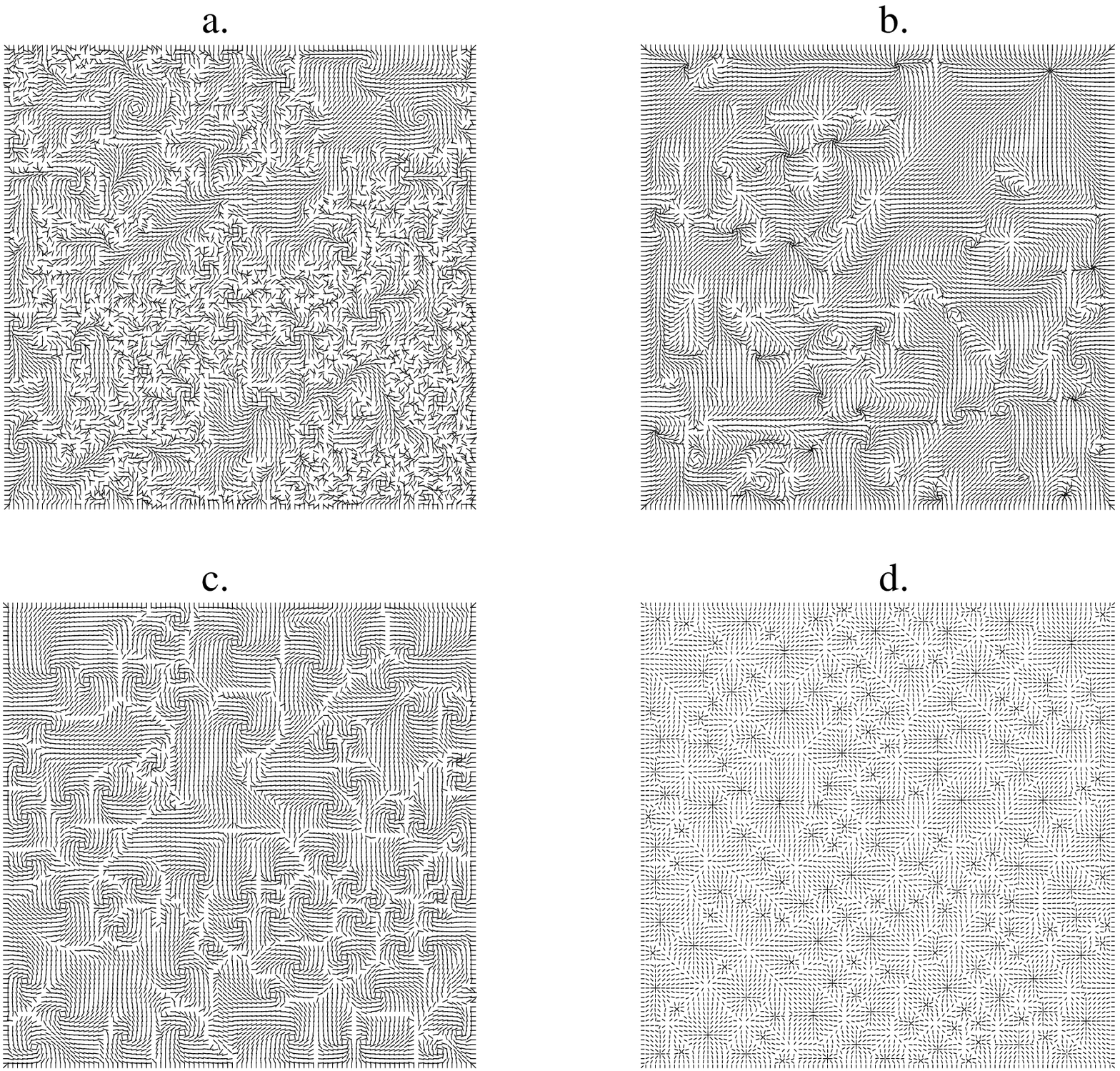}
\caption{Steady state configurations in our model 
at different parameter values (see text):
(a) Disordered phase
obtained at very low motor densities [$m = 0.005$
$\epsilon = 0.5$ and $S = 0$];
(b) Aster-vortex mixture obtained at
[$m = 0.01$, $\epsilon = 0.5$ and $S = 0$];
(c) Lattice of vortices at
[$m = 0.05$, $\epsilon = 5$ and $S=0.001$];
(d) Lattice of asters
obtained at [$m = 0.5$,
$\epsilon = 1$ and $S=1$]
}
\end{figure}

\newpage

\begin{figure}
\epsfbox{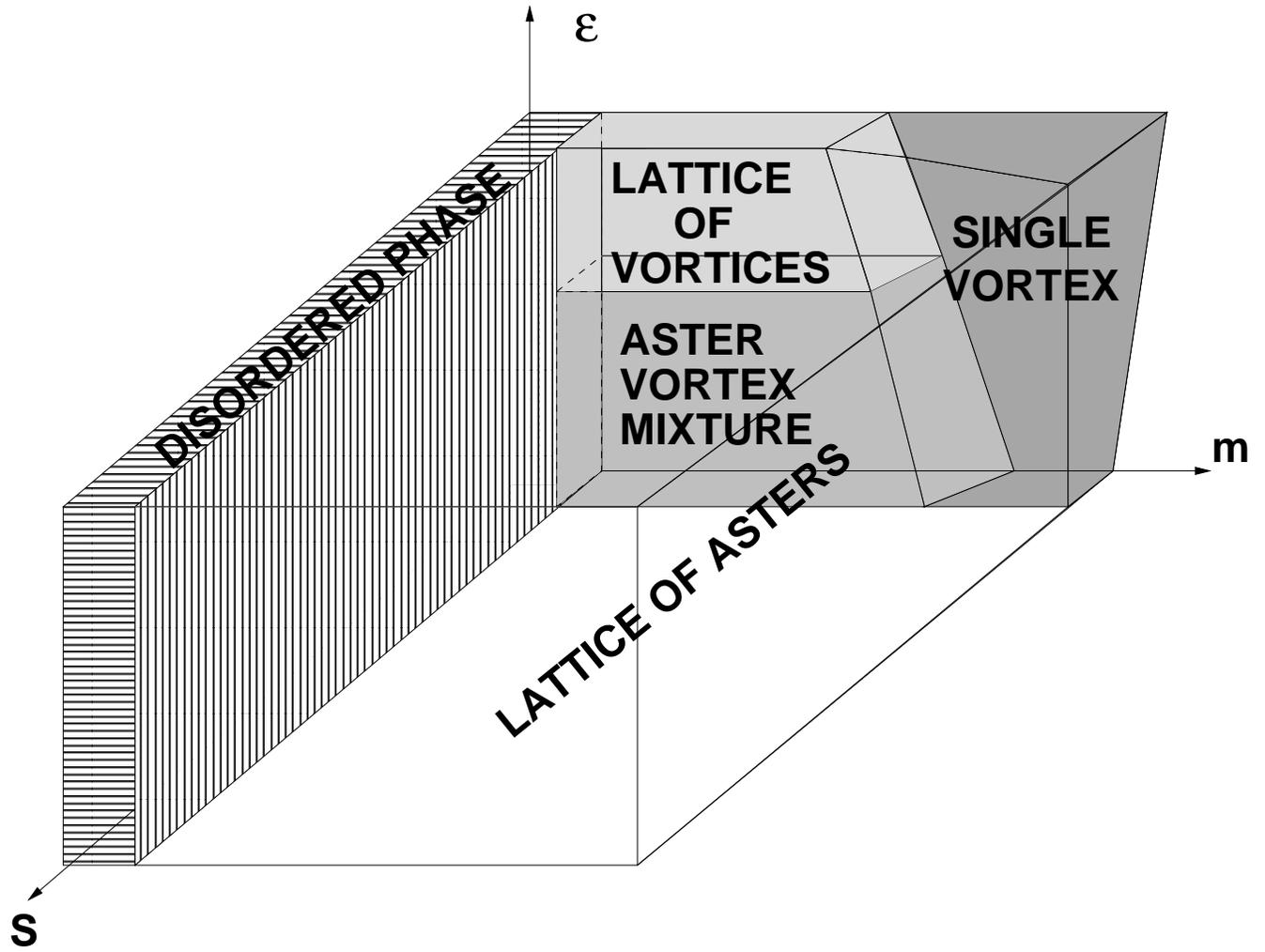}
\caption{Schematic phase diagram in the $\epsilon$, $m$ and $S$
plane, illustrating the regimes of parameter space in which
the phases of Fig. 1 are seen.
}
\end{figure}


\begin{thebibliography}{99}
\bibitem{notearray} Although we follow earlier authors
(see Refs.\cite{nedelecsurreymaggsleibler,surreynedelecleiblerkarsenti}),
in the use of  the term ``lattice'', we stress that these arrangements 
are not ordered in any regular manner.
\bibitem{notebundle} Disordered ``bundles'' 
are obtained at still larger $m$ in the
experiments of Ref.\cite{nedelecsurreymaggsleibler}, but are not
reproduced here.
\bibitem{notenoise} We add weak noise to the tubule equation to 
ensure that true steady states are reached and have checked,
varying both noise and initial conditions, that the patterns
we generate are robust.
\bibitem{notevortex} Our results for motor density profiles in
vortex configurations are the same as those of Ref.\cite{leekardar}.
\end{thebibliography}
\end{document}